\documentclass[12pt,twoside]{article}

\usepackage{amssymb}
\usepackage{amsmath}
\usepackage{latexsym}
\usepackage{longtable}
\usepackage{epsfig}
\usepackage{subfig}
\usepackage{bbold}

\begin{document}
\begin{titlepage}
\noindent
\underline{Contribution to the Proceedings of StatPhys24}
\begin{center}
\vspace*{2cm} {\Large {\bf The 1+1-dimensional Kardar-Parisi-Zhang
equation and its universality class\bigskip\bigskip\\}}
{\large{Tomohiro Sasamoto$^\star$ and
Herbert Spohn$^\dag$}}\bigskip\bigskip\\
  {$^\star$
  Chiba University, Chiba, Japan\\ e-mail:~{\tt sasamoto@math.s.chiba-u.ac.jp}}

 {$^\dag$  Zentrum Mathematik and Physik Department, TU M\"unchen,\\
 D-85747 Garching, Germany\\e-mail:~{\tt spohn@ma.tum.de}}\\
\end{center}
\vspace{5cm} \textbf{Abstract.} We explain the exact solution of the
1+1 dimensional Kardar-Parisi-Zhang equation with sharp wedge
initial conditions. Thereby it is confirmed that the continuum model
belongs to the KPZ universality class, not only as regards to
scaling exponents but also as regards to the full probability
distribution of the height in the long time limit.

\end{titlepage}


\section{Introduction}

Over the past year there has been significant advances in the
understanding of surface growth in 1+1 dimensions, both on the
experimental and theoretical side. The experiments on droplet growth
for a thin film of turbulent liquid crystal were presented at
StatPhys24 by K.A. Takeuchi \cite{TS10}. We will report here on the
theoretical findings, but use the occasion to put the results in a
wider context. We also take the liberty to comment on the
experimental situation.

In their seminal 1986 paper \cite{KPZ86} Kardar, Parisi, and Zhang
(KPZ) proposed an evolution equation for growing interfaces which in
1+1 dimensions, of interest here, reads
\begin{equation}\label{1}
\frac{\partial}{\partial
t}h=\tfrac{1}{2}\lambda\big(\frac{\partial}{\partial x}h\big)^2 +\nu
\frac{\partial^2}{\partial x^2}h+\sqrt{D}\eta\,.
\end{equation}
(\ref{1}) is a stochastic evolution equation for the height function
$h(x,t)$, hence has to be supplemented by the
appropriate initial conditions. The height $h$ depends on the
spatial location $x\in\mathbb{R}$ and on time $t\in\mathbb{R}_+$.
$\frac{1}{2}\lambda(\partial h/\partial x)^2$ is the nonlinear
growth velocity depending quadratically on the local slope. The
Laplacian smoothens the height profile with relaxation coefficient
$\nu>0$. $\eta(x,t)$ is space-time white noise, \textit{i.e.} $\eta$
is Gaussian with mean zero and covariance $\langle \eta(x,t)
\eta(x',t')\rangle=\delta(x-x') \delta(t-t')$. $\sqrt{D}$ is the
intensity of the noise and $\sqrt{D}\eta$ models the random nucleation events at
the interface. We consider only the infinite line without boundary conditions. Finite
geometry and possibly boundary sources are of physical interest. The 
corresponding large deviations have been studied by Brunet and 
Derrida \cite{BD00}.  In this article our main focus is 
on typical fluctuations. Their finite volume corrections have been  little explored so far. 

Already in their original contribution, KPZ noted that (\ref{1}) can
be transformed to a linear equation at the expense of turning the
additive noise $\eta$ into multiplicative noise. More precisely, one
introduces the Cole-Hopf transform
\begin{equation}\label{2}
Z(x,t)=\exp \big[(\lambda /2\nu) h(x,t)\big]\,.
\end{equation}
Then $Z$ satisfies
\begin{equation}\label{3}
\frac{\partial}{\partial t} Z(x,t)= \nu\frac{\partial^2}{\partial
x^2}Z(x,t)+(\lambda\sqrt{D}/2\nu) \eta(x,t) Z(x,t)\,.
\end{equation}
To ``solve'' (\ref{3}) we introduce the auxiliary standard Brownian motion
$b(t)$, $t\geq 0$, with variance $t$. Then, by the Feynman-Kac formula,
\begin{equation}\label{4}
Z(x,t)=\mathbb{E}_x \Big(\exp\big[\alpha \int^{2\nu t}_0 ds
\eta(b(s),s)\big] \exp\big[(\lambda/2\nu) h(b(2\nu t),0)\big]\Big)
\end{equation}
with $\alpha=(2\nu)^{-3/2} \lambda D^{1/2}$. To distinguish from the
white noise average denoted by $\langle\cdot\rangle$, the average
over the Brownian motion starting at $x$ is denoted here by
$\mathbb{E}_x(\cdot)$. $h(x,0)$ is the initial height profile.
According to (\ref{3}) the polymer has inverse stiffness $\nu$, which 
for notational convenience has been absorbed in the time scale of $b(t)$.

$Z(x,t)$ can be viewed as the random partition function of the
directed polymer $b(s)$, $0\leq s\leq 2\nu t$. The polymer is
subject to the random potential $\eta$ with strength $\alpha$. To
compute the potential energy  of the polymer one simply has to sum
$\eta$ along the location of the polymer chain.  Inverting (\ref{2}), the height is the
random free energy of the polymer. Thus the KPZ equation is
equivalent to a model in the equilibrium statistical mechanics of
disordered systems. Since the free energy valleys are in two-dimensional
space-time, the directed polymer is easier to visualize than the high-dimensional configuration space of
Ising spin glasses. 

We remark that the KPZ equation and the Cole-Hopf transformation
generalize in the obvious way from $1+1$ to $d+1$ dimensions.

There are books, comprehensive reviews,  and a large number of
original articles on growth models in the KPZ class
\cite{BS,K2,K,HZ}. Of course, we cannot provide here a comprehensive summary.
But very schematically, it seems reasonable to distinguish two in essence disjoint
lines of research.\smallskip\\
$\bullet$  Scaling theory, critical exponents, and
phase diagram in general dimension, starting with KPZ in 1986,\\
$\bullet$  exact probability density functions (pdf) for
growth models in $1+1$ dimensions, starting with Johansson \cite{Jo00b} in 2000.\medskip\\
We comment on a few general features.\medskip\\
\textit{Scaling theory, phase diagram}. Interface motion in the KPZ class
is characterized by local growth rules with single growth events
being statistically independent in space-time. There is no mass
transport along the interface. The respective bordering bulk phases
are statistically homogeneous and non-critical. As a rule, in the
course of time the interface develops a statistically self-similar
structure, when viewed on sufficiently coarse scales. The
self-similar structure has been amply confirmed, mostly by
Monte-Carlo simulations of various discretized models, for which there are
many options: space can be discretized to a lattice $\mathbb{Z}$,
the height may take only integer values, and one can implement
updates in discrete time steps. Slightly more distant from (\ref{1})
would be models as the off-lattice Eden growth model, where hard
balls of equal size are added at random one by one to the current
cluster of balls \cite{FA}. In fact, convincing numerical solutions
of the $1+1$ dimensional KPZ equation are only fairly recent
\cite{KS}. All these models constitute the KPZ universality class.
One expects that in the infrared scaling, \textit{i.e.} for large distances and long times, essentially all microscopic
details become irrelevant and universal statistical laws emerge.

Amongst the many advances the most important findings can be
summarized as follows.\smallskip\\
\textit{(i) Phase diagram.} For dimension $d\leq 2$ there is only a
strong coupling phase, excluding exceptional points where the
effective coupling constant vanishes. For $d>2$ there is a weak
coupling and strong coupling phase. In the weak coupling phase the
nonlinearity is irrelevant and fluctuations are according to the
linear $(\lambda=0)$ Edwards-Wilkinson equation. Over the last ten
years probabilists have proved in detail many of these claims. They
mostly consider a directed polymer, where $b(t)$ is replaced by a
discrete time random walk and $\eta(x,t)$ by a collection of
independent random potentials $V(j,t)$, $j\in\mathbb{Z}$,
$t\in\mathbb{Z}_+$. In particular, it is proved that in the weak
coupling regime the directed polymer has the statistics of a
Brownian motion with a diffusion constant computed from the annealed
version, see \cite{CY} for an overview and \cite{L09} for recent progress. In fact, the bounds on
the phase diagram are fairly sharp. From the perspective of spin
glasses a natural quantity to consider is the ratio $\langle
Z(x,t)^2\rangle/\langle Z(x,t)\rangle^2$. For $d>2$, at small
coupling the ratio stays bounded, while beyond a critical coupling,
$\lambda^{(2)}_\mathrm{c}$, it diverges exponentially. Based on
theoretical arguments and numerical simulations it has been claimed
that the true critical coupling
$\lambda_\mathrm{c}=\lambda^{(2)}_\mathrm{c}$ \cite{MG}. Rigorous
bounds establish $\lambda_\mathrm{c}>\lambda^{(2)}_\mathrm{c}$,
although with
a fairly small difference \cite{Bi,dH}.\smallskip\\
\textit{(ii) Critical exponents.} There is a static critical
exponent, $\chi$, which governs the height-height correlations in
the steady state at fixed slope and a dynamical critical exponent,
$z$. Both are related through
\begin{equation}\label{5}
z+\chi=2\,.
\end{equation}
For $d=1$, in the steady state of the KPZ equation the slope
$\partial h/\partial x$ is distributed according to spatial white
noise, hence $\chi=\frac{1}{2}$ and $z=3/2$. For $d>1$, 
extensive numerical simulations are available. In the analysis of the numerical
data one has to assume, implicitly or explicitly, a particular form
of the finite size scaling, for which no systematic theory is
available. This is one essential limiting factor in the attempt to
compare with theories making definite predictions for the value of
the
exponents.\smallskip\\
\textit{(iii) Nonuniversal constants.} All microscopic details of
either an experimental realization or of a particular theoretical
model are subsumed in three nonuniversal constants \cite{KMH92,
AF92}. One considers a particular direction along which the interface has
a non-random time-independent macroscopic slope $u$. Then the first coefficient is simply  
the steady state growth velocity $v_\infty$ reached for long times.
 The second one is the effective
coupling constant, defined by the Hessian $(\partial^2/\partial u_i
\partial u_j) v_\infty$. The KPZ theory assumes that the
eigenvalues of the Hessian are either all strictly positive or all
strictly negative. (If the eigenvalues have a different signature
the fluctuations will be Gaussian \cite{Wol91,BF08b}.) The third
coefficient, $A$, is the amplitude of the steady state correlation
function at slope $u$, i.e. $\langle(h(x)-h(x'))^2\rangle\cong
A|x-x'|^{2\chi}$ for \smallskip large $|x-x'|$. In general, all nonuniversal 
coefficients depend on the slope \smallskip $u$.

In the present context, the central unsolved problem is the upper critical
dimension. One
prediction, based on asymptotic expansions, is that $z$ increases with
$d$ and reaches the asymptotic value $2$ as $d\to\infty$. Other
field theory based approaches yield the upper critical dimension
$d=4$. For $d>4$ one finds $z=2$, like in the weak coupling phase,
but there are other statistical properties which still distinguish
between weak and strong coupling
phase. We refer to \cite{CD,L98,F06} and the recent letter by Canet
\textit{et al.} \cite{C10} with references to earlier literature.
\hspace{8 pt}$\diamondsuit$\medskip\\
\textit{ Exact solutions in 1+1 dimensions.} Ulam proposed the following
combinatorial problem: One considers a random permutation
$p(1)\ldots p(N)$ of $1\ldots N$. Any given permutation will have
increasing subsequences, \textit{e.g.} 6 4 1 5 2 3 has the
increasing subsequences 4 5, 1 2 3 and others. Ulam asked for  the
random length, $\ell_N$, of the longest increasing subsequence
(there could be several ones with the same length). One finds that
typically $\ell_N=\mathcal{O}(\sqrt{N})$ for large $N$. So the more
detailed question deals with the fluctuations relative to
$\sqrt{N}$. In a famous contribution, Baik, Deift, and Johansson
\cite{BDJ} established that the fluctuations are of order $N^{1/6}$.
While the exponent has been anticipated from numerical simulations,
the real surprise was that the pdf of the fluctuations turned out to
be identical to the Tracy-Widom distribution known from the Gaussian
unitary ensemble (GUE), thus establishing a completely unexpected
link to random matrix theory. Johansson \cite{Jo00b} extended these results to
the single step growth model (alias TASEP, the totally asymmetric
simple exclusion process). In this model space is
discrete, $j\in\mathbb{Z}$, and a height variable $h(j,t)$ takes
only integer values satisfying the single step constraint
$|h(j+1,t)-h(j,t)|=1$. In the stochastic update, independently at
each local minimum the height is increased by 2 with probability
$dt$ and stays put with probability $1-dt$. As initial condition we
set $h(j,0)=|j|$. Johansson proved that for the height at origin it
holds
\begin{equation}\label{6}
h(0,t)=\tfrac{1}{2} t + 2^{-1/3} t^{1/3} \xi_\mathrm{TW}
\end{equation}
for large $t$. Here $\xi_\mathrm{TW}$ is a GUE Tracy-Widom
distributed random variable, as will be discussed further below. As
predicted by the KPZ theory the fluctuations of the height are of
order $t^{1/3}$. Also, the numerical coefficients in (\ref{6}) are
the nonuniversal constants explained in \textit{(iii)} above.

In \cite{PrSp} it was shown that Ulam's problem is equivalent to the
polynuclear growth model (PNG) with droplet geometry, \textit{i.e.}
the height function has typically the shape of a semicircle. As a
surprising consequence, it was realized that while the size of the
fluctuations is always of order $t^{1/3}$, the statistics still
carries some information on the initial data
\cite{PrSp,BR00,BR99,BFPS06,BFS07}. \textit{E.g.}, in case of the
PNG model with flat initial conditions one finds that (\ref{6}) is
still valid, but the pdf of the random amplitude is given by the
Tracy-Widom distribution of the Gaussian orthogonal ensemble (GOE).

The discoveries from 2000 initiated ongoing activities. We refer to
a book  and reviews \cite{AGZ10,Spo05,FS10,KK10}, where more
details can be found.\medskip\hspace{8 pt}$\diamondsuit$

With the exact solutions obtained since 2000, one has now available
a more stringent test for a one-dimensional growth model to be in
the KPZ universality class: Not only must be the size of the
fluctuations of order $t^{1/3}$, but, for example in the droplet
geometry, the random amplitude must be GUE Tracy-Widom distributed.
For short, let us call this property the universal one-point pdf. One would expect
that any model in the KPZ  class shares the universal one-point pdf. In
fact, one should regard it as part of the definition for a model to
be in the KPZ universality class. So far the evidence comes from the
few models which allow for exact computations. One might have
guessed some activity from the side of Monte-Carlo simulations. In
the early period the statistical sampling was limited. But cumulants up to the fourth order
\cite{KMH92,AF92} and even the full height distribution function \cite{KMB91}
have been computed for various models, in retrospect achieving surprisingly good agreement with
the exact solution in 1+1 dimensions. More recently large scale simulations have been
carried out, \textit{e.g.} for off-lattice Eden models in the plane.
Exponents are measured with high precision \cite{FA}, but the full pdf
has not yet been determined.

The list of exactly soluble models is short.\medskip\\
$\bullet$ \textit{PNG model, TASEP with discrete and continuous time
update, PASEP.} These models are unified in the directed polymer picture.
They are zero temperature versions, \textit{i.e.} one studies the
fluctuations of the ground state energy of the directed polymer. In
the discrete time TASEP the directed polymer lives on the lattice
$\mathbb{Z}^2$ with the time direction along $(1,1)$. The droplet
geometry means that both endpoints of the polymer chain are fixed.
The random potential $V(i,j)$, $(i,j)\in\mathbb{Z}^2$, is
independent, identically distributed with a one-sided geometric
distribution, \textit{i.e.} $\mathrm{Prob}(\{V(i,j)=n\})=
(1-a) a^{|n|}$ for $n\leq 0$ and 0 otherwise with $0<a<1$. The PNG
model corresponds to the Poisson limit of rare events, $a\ll 1$,
while the continuous time
TASEP corresponds to the limit of a one-sided exponential.\medskip\\
$\bullet$ \textit{The partially asymmetric simple exclusion process
(PASEP)}. This is the single step model with a modified rule for
updates. As before local minima are increased by 2 at rate $p$ and,
in addition, local maxima are deceased by 2 with rate $q$, $q+p=1$.
The symmetric case $p=q=\frac{1}{2}$ has Gaussian fluctuations. For
$p> q$, Tracy and Widom \cite{TW2} prove the asymptotics of the form
(\ref{6}). \medskip\\
$\bullet$ \textit{KPZ equation}.
The KPZ equation (\ref{1}) has the static exponent
$\chi=\frac{1}{2}$, as a consequence of the exactly known steady
state, and hence the dynamic exponent $z=3/2$. To establish the 
universal one-point pdf requires a detailed computation, as will be
explained in the remainder of this contribution. For convenience we
set $\lambda>0$. The initial conditions are the sharp wedge $h(x,0)=
-|x|/\delta$ in the limit $\delta\to 0$. Then, for large $t$,
\begin{equation}\label{7}
h(0,t)=-\tfrac{1}{12}(\gamma_t)^3 +2 \log \alpha +\gamma_t
\xi_\mathrm{TW}\,,
\end{equation}
where $\gamma_t=(\alpha^4 \nu t)^{1/3}$. (\ref{7}) should be
compared with (\ref{6}). The prefactors are the nonuniversal
constants specific for the KPZ equation, while the logarithmic shift
has its origin in the slightly singular initial condition. In fact,
(\ref{7}) remains valid for every $t$ provided $\xi_\mathrm{TW}$ is
replaced by the random variable $\xi_t$. Its variance stays bounded
as a function of $t$ and its precise pdf is the main result of our
\medskip contribution. 

To close the introduction, we outline some experimental
activities.\medskip\\
(i) Ballistic deposition \cite{ToW94}. A rough surface is
generated by raining material onto a solid substrate. The randomness
results from the incident beam. One difficulty is the suppression of
surface diffusion. The incident velocity must be small to ensure a
proper local attachment. Also the intensity must be small so to
deposit on a relaxed surface. The KPZ fluctuations become visible
only in depositing many layers. The surface roughness is measured
through small angle scattering of X-rays. With these severe
constraints, at best a qualitative confirmation of the KPZ roughness
is obtained.\medskip\\
(ii) Smoldering paper \cite{MMMT05}. One fixes a piece of paper in a
frame and, after suitable chemical preparation, smolders it from the
bottom. It is argued that the one-dimensional combustion front is
governed by the KPZ equation. The random noise originates from the
intrinsic structure of the paper. Clearly, the geometry corresponds
to flat initial conditions and one would like to observe the 1/3
exponent and the GOE Tracy-Widom distribution. To discern such fine
statistical properties the experiment has to be repeated many times
with identical KPZ parameters, which is difficult to achieve. A total of 18, resp. 21, successful burns 
are recorded for horizontal, resp. tilted ignition. Using two fitting parameters for 
the probability density function, reasonable agreement with the theory
is obtained. \medskip\\
(iii) Facet edge fluctuations \cite{E06}. The experiment is
based on the observation that the 1+1 dimensional KPZ equation
governs also the thermal fluctuations of the edge of a crystal facet
(which is a one-dimensional structure). To measure directly the
spatial edge fluctuations is essentially impossible. However one can
observe the dynamical fluctuations of the facet edge which via
detailed balance are related to the static fluctuation exponent. The
experiment clearly rules out random walk like fluctuations of the
facet edge, characteristic for an isolated step, and favors the
reduced KPZ roughness. Qualitatively the reduction is caused by the
constraint through the neighboring step away from the facet and
by
entropic constraints for deviations into the facet.\medskip\\
(iv) Turbulent liquid crystal \cite{TS10}. One prepares a thin film
of liquid crystal with dimensions
$16\mathrm{mm}\times16\mathrm{mm}\times12\mu\mathrm{m}$. The film is
confined by two transparent electrodes and subject to oscillating
electric fields. A particular point,  temperature $25\; ^{\circ}\mathrm{ C}$ and voltage 26 V 
at 250 Hz, in the nonequilibrium phase
diagram is chosen with care, such that the so-called DSM 2 phase is
stable, while the DSM 1 phase is unstable. However the time for spontaneously
nucleating
the DSM 2 phase in the bulk DSM 1 phase is long compared with the
time of a single experimental run. Also unavoidable nucleations at
the border of the cell are fairly unfrequent. The system is prepared
in the homogeneous DSM 1 phase. A laser pulse plants a seed of the
DSM 2 phase which then grows within 35 sec to a droplet filling the
region seen by the camera. This is precisely the geometry used in the exact
solution of the KPZ equation. To have flat initial conditions the
laser pulse plants a line seed. The experiment is repeated of the
order of 1000 times. The DSM 1 phase is turbulent, hence good
mixing in the region away from the droplet is ensured. For a single
probe the physical parameters remain constant over long periods. By spatial
isotropy the average shape of the droplet is a disk which improves
substantially the analysis of the statistical data. First the
nonuniversal constants are determined with good precision. Thereby
the scales are fixed and no fitting parameters are used. The scaling
exponents, the GUE and GOE Tracy-Widom pdf, and the spatial
height-height correlations are measured. Very good agreement with
theory is achieved. Also finite time corrections are determined.


\section{The Cole-Hopf solution and its approximations}

As written, the KPZ equation is not well-defined mathematically. The
noise is singular and even if one takes the smoothening by the
Laplacian into account the solutions to the linear part of the
equation is rough. Thus the nonlinearity requires to multiply
pointwise two distributions which is an ambiguous mathematical
operation. The proper definition of solutions to the KPZ equation is
ongoing research. For our purposes the Cole-Hopf transform suffices.
Of course, it still reflects the difficulties mentioned: the action
integral is not well-defined for typical realizations of Brownian
motion and of white noise. Fortunately, as in other two-dimensional
field theories, an infinite energy renormalization suffices to make
sense out of (\ref{4}). Thus the strategy is to suitably approximate
$Z(x,t)$ and then to \textit{define} the solution to the KPZ
equation as
\begin{equation}\label{8}
h(x,t)=(2\nu/\lambda)\log Z(x,t)\,.
\end{equation}

We will study exclusively the sharp wedge initial conditions, which
under Cole-Hopf translates to
\begin{equation}\label{9}
Z(x,0)=\delta(x)\,.
\end{equation}

Currently there are five different approximation schemes, which all
yield the same limit stochastic process. We discuss each one
separately.\medskip\\
(A) Multiple It\^{o} integrals. We expand the exponential of
(\ref{4}) and average over Brownian motion which yields as $n$-th
coefficient
\begin{equation}\label{10}
\alpha^n  \int_{0\leq t_1\ldots\leq t_n\leq 2\nu t} dt_1\ldots dt_n
\int_{\mathbb{R}^n} dx_1\ldots dx_n \prod^n_{j=1}\eta(x_j,t_j) p_0
(x_1,t_1,\ldots,x_n,t_n,x,2\nu t)
\end{equation}
with $p_0$ the joint probability density for the Brownian motion to
start at 0, to be at $x_j$ at time $t_j$, $j=1,\ldots,n$, and to be
at $x$ at time $2\nu t$. We now integrate in $x$ and interpret the
remaining time integrations as multiple It\^{o} integral in the
forward discretization. Then by It\^{o}'s lemma
\begin{eqnarray}\label{11}
&&\hspace{-23pt}\langle Z(x,t)^2\rangle =\sum^\infty_{n=0} \alpha^{2n}
\int_{0\leq
t_1\ldots\leq t_n\leq 2\nu t}dt_1\ldots dt_n \nonumber\\
&&\hspace{40pt}\times\int_{\mathbb{R}^n} dx_1\ldots dx_n p_0
(x_1,t_1,\ldots,x_n,t_n,x,2\nu t)^2<\infty\,,
\end{eqnarray}
since the $n$-fold integral can be bounded by $C/(n!)^{1/2}$. Hence
$Z(x,t)$ is a well defined random variable. In fact, it can be shown
that $Z(x,t)>0$ and that $Z(x,t)$ is continuous in $x$ and in $t>0$
with probability one \cite{ACQ10}.\medskip\\
(B) Colored noise. One introduces the mollifier
$\varphi_\kappa(x)=\kappa\varphi(\kappa x)$, with $\varphi\geq 0$,
$\varphi$ of rapid decrease, $\varphi(x)=\varphi(-x)$, $\int dx
\varphi(x)=1$, and spatially smears the white noise to
$\eta_\kappa(x,t)=\int dx' \varphi_\kappa (x-x') \eta(x',t)$. The
Cole-Hopf transformation remains valid and the action is now a
properly defined integral. The white noise average over the
partition function is given by
\begin{eqnarray}\label{12}
&&\hspace{-26pt}\langle Z_\kappa(x,t)\rangle=\langle
\mathbb{E}_0\big(\exp\Big[\alpha\int^{2\nu t}_0
ds\eta_\kappa(b(s),s)\Big]\delta(b(2\nu t)-x)\big) \rangle \nonumber\\
&&\hspace{25pt} =p_0 (x,2\nu t) \exp[\tfrac{1}{2}\alpha^2
\varphi_\kappa \ast\varphi_\kappa(0) t]\,,
\end{eqnarray}
where $\ast$ denotes convolution and $p_0(x,t)=(2\pi t)^{-1/2}
\exp[-x^2/2t]$ is the transition probability of standard Brownian
motion. As $\kappa\to\infty$ one has $\eta_\kappa\to\eta$, but
$\langle Z_\kappa(x,t)\rangle$ diverges as
$\exp[\tfrac{1}{2}\alpha^2 \varphi \ast\varphi(0)\kappa t]$. On the
level of the KPZ equation this means that the average velocity of
the interface, $v_\kappa=\tfrac{1}{2}\alpha^2 \varphi_\kappa
\ast\varphi_\kappa(0)$, diverges linearly in $\kappa$. But it can be
proved that in the frame moving with constant velocity $v_\kappa$ in
the $h$-direction one nevertheless has a well-defined limit as
$\kappa\to\infty$. In other words
\begin{equation}\label{13}
\lim_{\kappa\to\infty} p_0(x,t) Z_\kappa(x,t)/\langle
Z_\kappa(x,t)\rangle=Z(x,t)
\end{equation}
with $Z(x,t)$ as defined in item (A).\medskip\\
(C) Lattice directed polymer. The integral (\ref{4}) defining the
random partition function is discretized. One replaces the Brownian
motion $b(t)$ by the random walk $\omega$ on the lattice
$\mathbb{Z}^2$. $\omega$ starts at $(0,0)$ and moves with
probability $\frac{1}{2}$ either up or right. At each site of the
lattice there is independently a unit Gaussian random potential
$\eta(i,j)$. Then the discrete approximation to (\ref{4}) reads
\begin{equation}\label{14}
Z_N (x,t)=\sum_{\omega:(0,0)\rightsquigarrow((t+x)N,(t-x)N)} 2^{-t
N} e^{-\beta E(\omega)}\,,
\end{equation}
where the sum is only over paths with endpoint $((t+x)N,(t-x)N)$,
$|x|<t$, and the energy of the walk $\omega$ is defined through
\begin{equation}\label{15}
E(\omega)=\sum_{(i,j)\in\omega} \eta(i,j)\,.
\end{equation}
As proved in \cite{AKQ}, in the limit $N\to\infty$, $\beta^4
N=\alpha$ fixed,
\begin{equation}\label{16}
\lim_{N\to\infty} Z_N(x,t)= Z(x,t)
\end{equation}
as a stochastic process and $Z(x,t)$ as in item (A) with parameters
$\nu=\frac{1}{2}$ and
$D=1$.\medskip\\
(D) The attractive $\delta$-Bose gas. For integer moments of the
partition function with colored noise as in (B) one can carry out
the Gaussian average over $\eta$ with the result
\begin{equation}\label{17}
\langle Z_\kappa(x,t)^n\rangle=\langle 0|e^{-2\nu t
H_n(\kappa)}|x\rangle\,.
\end{equation}
On the right one has a matrix element of a quantum propagator for
$n$ particles on $\mathbb{R}$. More explicitly
\begin{equation}\label{18}
H_n(\kappa)=-\sum^n_{j=1} \tfrac{1}{2}\frac{\partial^2}{\partial
x^2_j}-\tfrac{1}{2} \alpha^2 \sum^n_{i,j=1} \varphi_\kappa
\ast\varphi_\kappa(x_i-x_j)
\end{equation}
with $x_j\in\mathbb{R}$ the position of the $j$-th quantum particle
and $|x\rangle$ denoting the quantum state where all $n$ particles
are at $x$. As $\kappa\to\infty$, $Z_\kappa(x,t)/\langle
Z_\kappa(x,t)\rangle$ converges to $Z(x,t)/p_0(x,t)$ with $Z(x,t)$
as in (A), while on the right one finds the hamiltonian for $n$
quantum particles with an attractive $\delta$-potential and the
self-term omitted. Therefore
\begin{equation}\label{19}
\langle Z(x,t)^n\rangle=\langle 0|e^{-2\nu t H_n}|x\rangle\,,
\end{equation}
\begin{equation}\label{20}
H_n=-\sum^n_{j=1} \tfrac{1}{2}\frac{\partial^2}{\partial
x^2_j}-\tfrac{1}{2} \alpha^2 \sum^n_{i\neq j=1} \delta(x_i-x_j)\,.
\end{equation}
While (\ref{19}), together with (\ref{20}), is a correct identity,
it cannot be used to define the pdf of $Z(x,t)$, since $\log\langle
Z(x,t)^n\rangle\cong n^3$ and the moment problem defined by
(\ref{19}) has many solutions \cite{T07}.\medskip\\
(E) Weak ASEP. We consider the single step model with asymmetry
$q$, $q>\frac{1}{2}$. The PASEP occupation variables are denoted by
$\eta_j(t)$, $j\in \mathbb{Z}$, $\eta_j(t)=0,1$. The initial height
profile is $h(j,0)=-|j|$. For this initial condition the height at
time $t$ is given by
\begin{equation}\label{21}
h(j,t)=-2\sum^j_{\ell=-\infty} \eta_\ell (t)+j\,.
\end{equation}
We have seen that for the directed polymer the KPZ equation
corresponds to a weak noise approximation. In a similar spirit, the
PASEP should approximate the KPZ for weak asymmetry. The asymmetry
must be carefully chosen. If it is too weak, one arrives at the
Gaussian theory with $\lambda=0$ and, if the asymmetry is too strong
one misses the approximation through a stochastic partial
differential equation altogether. The correct choice is as follows:
first we scale space-time diffusively, \textit{i.e.}
$j=\lfloor\varepsilon^{-1} x\rfloor$ with $\lfloor\cdot\rfloor$
denoting integer part and time as $\varepsilon^{-2}t$,
$t=\mathcal{O}(1)$. $0<\varepsilon\ll 1$ is the scaling parameter.
With this choice we set $q+p=1$ and $q-p=\beta\sqrt{\varepsilon}$,
$\beta>0$. To recall the $\varepsilon$-dependence of $q$, we write
$h^\varepsilon(j,t)$ instead of $h(j,t)$ in the definition
(\ref{21}). As established in \cite{BG,ACQ10}, as convergence of
stochastic processes it holds
\begin{equation}\label{22}
\lim_{\varepsilon\to 0} \sqrt{\varepsilon}\beta h^\varepsilon
(\lfloor \varepsilon^{-1} x\rfloor, \varepsilon^{-2} t)+
\tfrac{1}{2}\beta^2 t \varepsilon^{-1} -\tfrac {1}{24}\beta^4 t-\log
(2\sqrt{\varepsilon}/\beta)=\beta h(x,t)\,.
\end{equation}
Here $h$ on the right side is defined by (\ref{8}) with $Z(x,t)$ as
in item (A) and parameters $\nu=\frac{1}{2}$, $D=\frac{1}{4}$, and
$\lambda=\beta$.\medskip

The reader might ask, why we make such an extensive list. Firstly,
the items provide a better understanding under which physical
conditions the KPZ equation is a valid approximation. Secondly, the
list ensures that the natural but somewhat formal procedure in (A)
and (B) properly captures our understanding based on discrete
models. Thirdly, and perhaps most importantly, the various
approximations do not only give mathematical sense to the KPZ
equation but also provide a tool by which, at least in principle,
some properties of the KPZ equation can be computed. In this
respect, (A) can be used for a short time expansion. (C) has not
been of help, yet. But a related discretization, where the
directed polymer is placed on $\mathbb{Z}\times\mathbb{R}$, has been
used to compute the free energy \cite{BBO06}. The identity in (D) is
the gateway to the replica method as used for spin glasses and other disordered 
systems with the special
feature that the matrix element on the right side can be analysed
through the Bethe ansatz. While the replicas forces one to work
with divergent series, the method has turned out to be a powerful
tool. Our exact solution is based on the approach (E). But at an
intermediate stage we have to rely on a deep theory for the PASEP
developed by Tracy and Widom in recent years
\cite{TW0,TW1,TW2}.


\section{Exact solution of the KPZ equation with sharp wedge initial conditions}

The pdf of the random free energy $\log Z(x,t)$ with $Z(x,t)$ as in
Section 2 has been computed for every $x,t$. We use a Fredholm determinant
formula by Tracy and Widom \cite{TW2}, which in case of 0\,-1 step initial conditions
provides the probability distribution for the position of the $m$-th particle at time $t$ and thereby
the probability distribution of $h^{\varepsilon}$ of (\ref{22}). While this is a convenient starting point,
to actually establish the limit in (\ref{22}) still requires an intricate asymptotic analysis.

To explain the structure of the exact solution we
first recall the pdf for the Tracy-Widom distributed random variable
$\xi_\mathrm{TW}$ of (\ref{6}). Like the pdf of $h(x,t)$, it is
defined in terms of a Fredholm determinant of a symmetric integral
operator with kernel over $\mathbb{R}\times\mathbb{R}$. We introduce
the Airy kernel
\begin{equation}\label{23}
K_\mathrm{Ai}(x,y)=\int^\infty_0 d\lambda \mathrm{Ai}(x+\lambda)
\mathrm{Ai}(y+\lambda)\,,
\end{equation}
where Ai is the standard Airy function. The corresponding operator
acting on $L^2(\mathbb{R},dx)$ is denoted by $K_\mathrm{Ai}$.
$K_\mathrm{Ai}$ is a symmetric projection. We also introduce the
projection onto the interval $[s,\infty)$, denoted by $P_s$. $P_s
K_\mathrm{Ai} P_s$ is trace class for every $s>-\infty$. Then
\begin{equation}\label{24}
 \mathrm{Prob}(\xi_\mathrm{TW}\leq s)= \det (1-P_s
 K_\mathrm{Ai}P_s)\,.
 \end{equation}
The Fredholm determinant on the right can be defined through the
eigenvalues $\lambda_j(s)$, $j=1,2,\ldots$, of $P_s K_\mathrm{Ai}
P_s$ as
\begin{equation}\label{25}
\det (1-P_s K_\mathrm{Ai} P_s)=\prod^\infty_{j=1}
(1-\lambda_j(s))\,.
\end{equation}
In fact, it is more appropriate to think of $P_s K_\mathrm{Ai} P_s$
as a large matrix by evaluating as
\begin{equation}\label{26}
A^{(N)}_{ij}= \chi_s (x_i)
K_{\mathrm{Ai}}(x_i,x_j)\chi_s(x_j)\,,\quad i,j=1,\ldots,N\,,
\end{equation}
with $\{x_j,j=1,\ldots,N\}$ suitably chosen, \textit{e.g.} equally
spaced, base points, and $\chi_s$ the indicator function of the
interval $[s,\infty)$. Then
\begin{equation}\label{27}
\det(1-P_s K_\mathrm{Ai} P_s)\cong \det (1-A^{(N)})\,,
\end{equation}
for large $N$. The optimal choice of base points and error estimates
are discussed in \cite{Bo}.

For the solution of the KPZ equation we obtain
\begin{equation}\label{28}
(\lambda/2\nu) h(x,t)=-(x^2/4\nu t)-\tfrac{1}{12}(\gamma_t)^3 +2\log
\alpha +\gamma_t \xi_t\,,
\end{equation}
with $\gamma_t=(\alpha^4 \nu t)^{1/3}$, $\alpha=(2\nu)^{-3/2}
\lambda D^{1/2}$. We note the self-similar flattening of the
deterministic droplet shape proportional to $t$ and a uniform shift,
also proportional to $t$. The relative fluctuations are of order
$t^{1/3}$ with a random amplitude $\xi_t$ which is of order 1 but
has a pdf changing in time. In (\ref{7}) we anticipated the special
case $x=0$ and $t$ large. The pdf of $\xi_t$ is given by
\begin{eqnarray}\label{29}
&&\hspace{-40pt} \rho_t (s)=\int^\infty _{-\infty} du
\gamma_t e^{\gamma_t(s-u)} \exp[-e^{\gamma_t(s-u)}]\nonumber\\
&&\hspace{25pt}\times \big(\det(1-P_u(B_t-P_\mathrm{Ai})P_u)-\det
(1-P_u B_t P_u)\big)\,.
\end{eqnarray}
Here $B_t$ has the
kernel
\begin{equation}\label{30}
B_t(x,y)= \int^\infty_{-\infty} dw (1-e^{-\gamma_t w})^{-1}
\mathrm{Ai}(x+w) \mathrm{Ai}(y+w)\,.
\end{equation}
and $P_\mathrm{Ai}$ is an one-dimensional projection  with
integral kernel $\mathrm{Ai}(x)\mathrm{Ai}(y)$. The first factor under the integral in (\ref{29}) is the pdf of the
Gumbel distribution known from extreme statistics. (\ref{29}) looks
like the convolution of two pdfs. However from numerical simulations
we know that the second factor, while normalized, may take negative
values.

Equipped with the exact solution (\ref{29}), we can check whether its long time 
is given by the GUE Tracy-Widom distribution, to say whether
$\xi_t\to\xi_\mathrm{TW}$ as $t\to \infty$. Clearly, for
$t\to\infty$, the first factor of (\ref{29}) converges to
$\delta(s-u)$ and the second factor to
\begin{equation}\label{31}
\det(1-P_u(K_\mathrm{Ai}-P_\mathrm{Ai})P_u)-\det (1-P_u
K_\mathrm{Ai} P_u)= \frac{d}{du}\det (1-P_u K_\mathrm{Ai} P_u)\,,
\end{equation}
the latter inequality following from identities proved in
\cite{TW94}. We conclude that the universal one-point pdf is established. For very short times $\xi_t$ has a Gaussian
distribution of
width $t^{1/4}$ \cite{ACQ10}. Hence our solution to the KPZ equation describes the 
crossover from a Gaussian at short times to the GUE
 Tracy-Widom at long times. Plots of the pdfs covering mostly late times can be found in \cite{SS2}.
\medskip\\
{\textit{Note.}} The story behind the exact solution
(\ref{28}), (\ref{29}) is slightly convoluted. The formula was
derived in the fall 2009 by Amir, Corwin, and Quastel
\cite{ACQ10} and  independently by us \cite{SS10,SS2,SS1}. All papers use as their starting point
a recent Fredholm determinant formula by
Tracy and Widom for the PASEP with 0\,-1 step initial conditions valid
for any $q > 1/2$, in particular for $q=\frac{1}{2}+2\beta
\sqrt{\varepsilon}$ \cite{TW2}. Simultaneously the replica method was persued.
Dotsenko and Klumov \cite{D1,D2} provide a detailed analysis of the
eigenfunctions of the attractive $\delta$-Bose gas on the line. With the hindsight from the
exact solution, this
leads to 
\begin{eqnarray}\label{32}
&&\hspace{0pt}\big\langle \exp{\big[-\exp(-\lambda + \log Z(x,t)+(x^2/4\nu t)+\tfrac{1}{12}(\gamma_t)^3  - 2\log
\alpha) \big]}  \big\rangle\nonumber\\[1ex]
&&\hspace{100pt}= \det (1-K_{\lambda,t})
\end{eqnarray}
for all real $\lambda$ and $t > 0$, expressing the generating function by a Fredholm determinant
 \cite{CLR,D}. The operator $K_{\lambda,t}$ has the kernel
\begin{equation}\label{33}
K_{\lambda,t}(x,y) =  (1+e^{(\gamma_t x - \lambda)})^{-1}e^{(\gamma_t x - \lambda)}
K_{\mathrm{Ai}}(x,y).
\end{equation}
Substituting $\lambda$ by $\gamma_t\lambda$  in (\ref{32}) and taking the limit $t \to \infty$ on both sides
yields indeed the GUE Tracy-Widom distribution function for the rescaled height.
 Calabrese, Le
Doussal, and Rosso \cite{CLR} establish that if the average in (\ref{32}) is computed by using (\ref{29})
one indeed arrives at $\det (1-K_{\lambda,t})$. \medskip\hspace{8 pt}$\diamondsuit$

 Equipped with such an input the universality of the KPZ equation can be discussed with more precision than before. One aspect concerns how well a microscopic system is approximated by the KPZ equation, which to be meaningful requires a tunable asymmetry. For example, the polynuclear growth model and also the liquid crystal of the experiment would have no such tunable parameter. But if available, then, roughly speaking, the description by the KPZ equation becomes valid in the limit of weak asymmetry and correspondingly long times. An example
in case is the 2D Ising model with Glauber dynamics at low temperatures with the $+\, -$ interface oriented along the (1,1) direction. The asymmetry parameter is the external magnetic field. For weak fields, which is needed anyhow in order to suppress nucleation in the bulk, the motion of the interface is well described by the KPZ equation.

As a second aspect we note that the exact solution of the KPZ equation
provides us with qualitative information beyond the universal features. To illustrate we discuss the   
example of the approach to the Tracy-Widom distribution. In the experiment \cite{TS10},
and also for the TASEP with 0\,-1 step initial conditions, one finds that the mean is the slowest mode, while higher cumulants decay rapidly to their Tracy-Widom value. Schematically, for large $t$ we write
\begin{equation}\label{34}
\langle h(0,t) \rangle = v_\infty t + a_1 t^{1/3}(\langle\xi_{\mathrm{TW}}\rangle + c_0t ^{-1/3}) + \mathcal{O}(t^{-1/3}).
\end{equation}
Here $v_\infty$ and $a_1$ are model dependent parameters and we discuss the coefficient $c_0$. $c_0 > 0$ for the experiment and TASEP. Relative to $\langle\xi_{\mathrm{TW}}\rangle$ the mean decays as
$t^{-1/3}$ with a positive amplitude. On the other hand one can expand (\ref{29}) in $1/t$.
The expansion is dominated by the Gumbel distribution, which produces a shift by $t^{-1/3}$
to the  left. Thus the exponent comes out correctly, but the amplitude and the full first order finite time correction are model dependent. Since 
 the KPZ equation corresponds to weak asymmetry, the PASEP 
 with $q$ close to $1/2$ should have an approach also from the 
 left. This is indeed confirmed by Monte Carlo simulations, which show
 a sign change of $c_0$ at $q \approx 0.7$. 
  
 \section{Conclusions and outlook}
 
By our results we have added one prominent member to the KPZ 
universality class. Currently  more  
universal pdfs have been computed for lattice models. For curved 
geometry and for flat initial conditions one knows not only the pdf for a 
single location but also all multi-point pdfs at the same long time, see
\cite{F10} for the current status. In addition 
the case of stationary initial conditions has been studied.
For the KPZ equation this would mean to choose an initial height such 
that $\partial h(x,0)/\partial x $ is white noise in $x$.
One has to see whether the KPZ equation will catch up with these results in the future.

From the point of view of the directed polymer, the continuum KPZ version
is the only finite temperature model for which universal pdfs have been 
obtained so far. It would be of interest to understand how a comparable result can 
be accomplished for lattice models. \bigskip\\
\textbf{Acknowledgements}. We are grateful to Patrik Ferrari and Joachim Krug for most useful comments.

\end{document}